\pdfoutput=1
\documentclass[a4paper, oneside, twocolumn, notitlepage, 10pt]{extarticle_ecoc}
\usepackage{ecoc}
\usepackage{physics}
\usepackage{xcolor}
\usepackage[normalem]{ulem}
\usepackage{setspace}
\usepackage{xurl}
\usepackage{graphicx}
\usepackage{microtype}
\usepackage{booktabs}
\hypersetup{
    pdftitle={GsOQDC: A GUI-Driven Interactive Framework for End-to-End Simulation of Optical Quantum Data Centers},
    pdfauthor={Seyed Navid Elyasi, Sima Bahrani, Rui Wang, Dimitra Simeonidou, Paolo Monti, Rui Lin},
    pdfsubject={ECOC 2026 Demo Proposal},
    pdfkeywords={Distributed Quantum Computing, Optical Quantum Data Centers, Quantum Networking}
}

\begin{document}
\selectlanguage{english}

\title{GsOQDC: A GUI-Driven Interactive Framework for End-to-End Simulation of Optical Quantum Data Centers}

\author{
    Seyed Navid Elyasi\textsuperscript{(1)}, Sima Bahrani\textsuperscript{(2)},
    Rui Wang\textsuperscript{(2)},
    Dimitra Simeonidou\textsuperscript{(2)}, 
    Paolo Monti\textsuperscript{(1)},
    Rui Lin\textsuperscript{(1)}}

\maketitle                  % Create title and author

%------------------------------------------ Description of Authors ----------------------------------------------%

\begin{strip}
    \begin{author_descr}

        \textsuperscript{(1)} Department of Electrical Engineering, Chalmers University of Technology, Gothenburg, Sweden
        
        \textsuperscript{(2)} Smart Internet Lab, School of Electrical, Electronic, and Mechanical Engineering, University of Bristol, Bristol
        \textcolor{blue}
        {\uline{elyasi@chalmers.se}}

    \end{author_descr}
\end{strip}

\renewcommand\footnotemark{}
\renewcommand\footnoterule{}
\begin{figure*}[b!]
    \centering
\includegraphics[width=\textwidth]{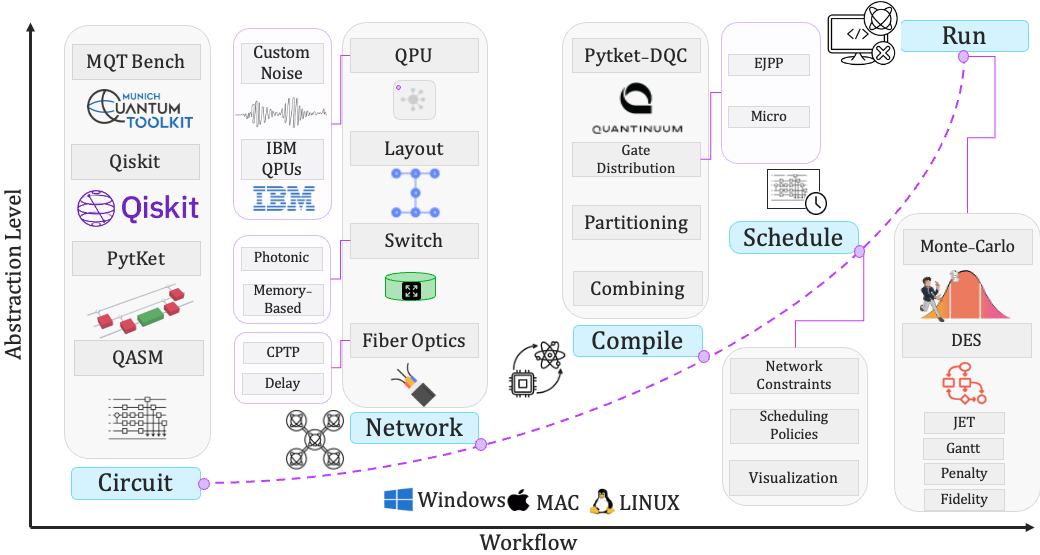}
    \caption{GsOQDC workflow and software architecture. The platform integrates the end-to-end DQC workflow, enabling users to progress from quantum-circuit design and optical-network configuration to distributed compilation, scheduling, execution, and performance evaluation within a single graphical environment.}
    \label{fig:workflow}
\end{figure*}
\begin{strip}
    \begin{ecoc_abstract}
We present GsOQDC, an open-source graphical framework integrating optical-network design, distributed quantum-circuit compilation, scheduling, and DES-based remote-gate simulation, enabling end-to-end cross-layer evaluation of entanglement-resource dynamics and system-level performance in Optical Quantum Data Centers \copyright 2026 The Author(s)
    \end{ecoc_abstract}
\end{strip}

%------------------------------------------------------------------------
\section{Review}
 
Quantum Data Centers (QDCs), in which multiple Quantum Processing Units (QPUs) are interconnected through optical networks, are emerging as a promising architecture for scaling quantum computing beyond the limits of monolithic devices \cite{elyasi2025framework,main2025,caleffi2024}. Recent experimental demonstrations of distributed quantum computation (DQC) across optically linked quantum modules \cite{main2025,Zhou2026PhotonicLink,Krutyanskiy2023} have renewed interest in DQC, where quantum circuits are partitioned across multiple QPUs and coordinated through entanglement distribution over an optical channel\cite{main2025, mirhosseini2020,ionq2022, main2025}.

However, the software ecosystem for QDC research remains fragmented. Quantum software frameworks support circuit design and compilation\cite{tket,qiskit2024}, distributed compilers address circuit partitioning\cite{andres2024}, and network simulators \cite{cisco_packet_tracer}model communication infrastructures independently. As a result, researchers lack an integrated environment for evaluating the end-to-end behavior of distributed quantum applications and the interactions between quantum-computing and optical-network resources.

We present \textbf{GsOQDC (\textit{GUI-Driven Simulator for Optical Quantum Data Centers})}, an open-source graphical framework for the design, simulation, and evaluation of Optical Quantum Data Centers. As shown in Fig.~\ref{fig:workflow}, GsOQDC unifies the end-to-end DQC workflow, Circuit, Network, Compile, Schedule, and Run, within a single interactive environment. The platform integrates quantum-circuit modelling, optically switched QPU interconnection, network-aware distributed compilation, gate-level circuit scheduling, and a discrete-event simulation (DES) engine, enabling end-to-end cross-layer performance evaluation. Through an intuitive graphical interface, users can interactively construct QDC architectures, generate distributed workloads, visualize resource utilization, and investigate system-level trade-offs across the application, system, and network layers. The demonstration showcases an end-to-end DQC workflow, enabling rapid design-space exploration, reproducible experimentation, and quantitative analysis of how optical-network characteristics influence application-level performance in future Optical QDCs.

%------------------------------------------------------------------------
\section{Novelty}
\begin{figure*}[t!]
    \centering
    \includegraphics[width=\textwidth]{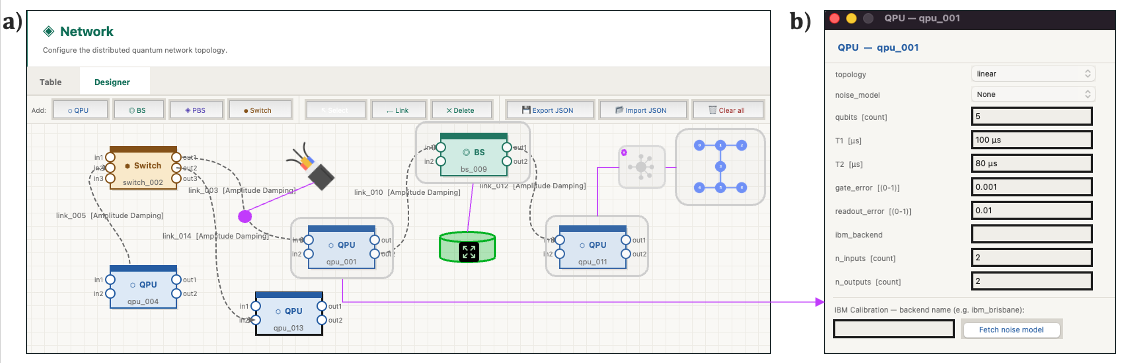}
    \caption{GsOQDC Network Designer. (a) Interactive topology editor for constructing Optical Quantum Data Center architectures using QPUs, optical components, and communication links. (b) Component configuration panel, accessible via right-click, allowing users to modify hardware and noise parameters, including qubit properties, gate errors, coherence times, readout errors, and device-specific calibration data. Together, these capabilities enable realistic, cross-layer modeling of distributed quantum-computing infrastructures.}
    \label{fig:designer}
\end{figure*}
 
GsOQDC provides an integrated environment for the design and evaluation of distributed quantum applications over Optical Quantum Data Centers (OQDC). As illustrated in Fig.\ref{fig:workflow}, the platform unifies the key components of a distributed quantum computing stack, spanning circuit specification, compilation, partitioning, scheduling, and network-level simulation, allowing researchers to move seamlessly from application specification to system-level analysis.

A key capability of GsOQDC is its network-centric design environment (Fig.~\ref{fig:designer}), which elevates the optical infrastructure from a hidden communication abstraction to a configurable design object. Existing quantum-network simulators such as NetSquid \cite{coopmans2021netsquid} and QuNetSim \cite{diadamo2021qunetsim} provide detailed protocol-level modeling of quantum communication systems but do not support distributed quantum-circuit compilation, scheduling, or system-level execution analysis. 

Conversely, DQC frameworks and partitioning tools focus on circuit decomposition and resource allocation while typically assuming simplified or abstract communication models \cite{zhang2024gantt, simdisq2025, andres2024}. GsOQDC bridges this gap by integrating QPU placement, optical switching architectures, quantum memories, optical links, distributed compilation, scheduling, and system-level evaluation within a single graphical environment. This enables direct investigation of how network architecture and physical-layer impairments impact system-level metrics such as fidelity, communication overhead, entanglement-resource utilization, and Job Execution Time (JET).

\begin{figure*}[t!]
    \centering
    \includegraphics[width=\textwidth]{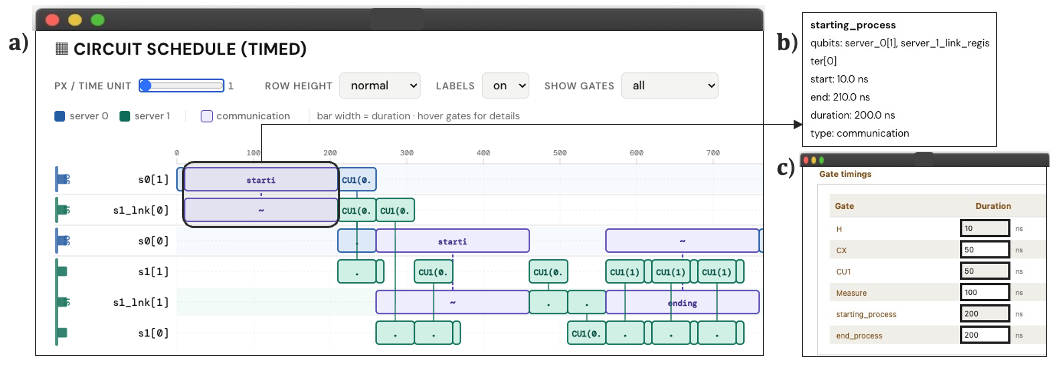}
    \caption{Distributed-workload scheduling in GsOQDC. (a) Interactive Gantt-chart visualization showing local operations, remote-gate execution, and communication events across multiple QPUs. (b) Contextual information panel displayed when hovering over a scheduled operation, providing timing, duration, resource, and process details. (c) User-configurable gate and process timing parameters.}
    \label{fig:schedule}
\end{figure*}

The platform further integrates distributed compilation and scheduling into a unified workflow. Compiled circuits are partitioned across multiple QPUs using configurable partitioning strategies, while interactive scheduling views as Gannt representation(Fig.~\ref{fig:schedule}) expose the temporal relationship between local operations, remote-gate execution, and communication events. By visualizing resource contention and communication overheads, GsOQDC provides insight into performance bottlenecks that are typically hidden within standalone compilation frameworks. In particular, the platform enables the design and evaluation of scheduling strategies for remote gates, a critical consideration given that remote gate execution times are significantly longer than local gates, directly affecting qubit decoherence and overall circuit fidelity.

The final stage employs a custom DES and Monte Carlo engine to model the execution of remote quantum operations across distributed QPUs. The framework captures the end-to-end lifecycle of remote-gate execution, including entanglement generation, distribution, storage, consumption, and communication-induced delays. This enables direct evaluation of entanglement fidelity, JET, entanglement-resource utilization, and penalty delays caused by entanglement unavailability.

To the best of our knowledge, GsOQDC is the first graphical platform that enables end-to-end design-space exploration of distributed quantum applications while jointly considering optical-network behavior, entanglement-resource dynamics, and system-level performance.Unlike existing research tools that typically require command-line interaction and  programming expertise, GsOQDC provides a user-friendly graphical environment that lowers the barrier to entry for DQC research. As shown in Fig.\ref{fig:workflow}, the platform is designed as a standalone cross-platform application and can be readily deployed on conventional Windows, macOS, and Linux systems, facilitating accessible and reproducible experimentation across the quantum networking and optical communications communities.

%------------------------------------------------------------------------
\section{ECOC Relevance}
 
Optical networking is emerging as a key enabler of scalable DQC, where optical interconnects distribute entanglement between physically separated QPUs. In such architectures, network-level decisions, including topology design, optical switching, entanglement generation policies, communication latency, and optical impairments, directly influence application-level performance. Understanding these interactions is therefore essential for the design of future Optical Quantum Data Centers and quantum-enabled network infrastructures.

GsOQDC addresses this challenge by providing an end-to-end framework for the design, scheduling, and evaluation of DQC  applications over configurable optical-network architectures. By enabling cross-layer analysis across the application, compilation, scheduling, and network stack, the platform allows to quantify the impact of optical network characteristics on metrics such as entanglement fidelity, JET, inter-QPU communication overhead, and entanglement-resource utilization. The demonstration highlights a novel and timely application of optical communication technologies, aligning with ECOC's interests in advanced optical networks, data-center interconnects, and emerging quantum networking systems.
%------------------------------------------------------------------------
\section{Demo Implementation}
The demonstration will be conducted using the GsOQDC graphical platform running on a standard workstation and will showcase the end to end lifecycle of a DQC executed over a QDC. Attendees will be guided through the workflow illustrated in Fig.~\ref{fig:workflow}, beginning with quantum-circuit specification and progressing through optical-network design, distributed compilation, scheduling, and system-level performance evaluation.

Using the interactive Network Designer (Fig.~\ref{fig:designer}), attendees can construct QDC topologies by placing QPUs, optical switching elements, entanglement-generation resources, quantum memories, and optical links. Hardware and noise parameters can be configured through component-specific dialogs, enabling rapid exploration of different architectural and physical-layer assumptions. The configured infrastructure is then used to compile quantum circuits across multiple QPUs using distributed partitioning algorithms, exposing communication requirements, remote-gate dependencies, and entanglement-resource demands.

The resulting distributed workload is transformed into an executable schedule and visualized through interactive Gantt charts (Fig.~\ref{fig:schedule}). Attendees can inspect individual operations, communication events, and resource allocations while observing how network constraints and scheduling policies influence execution timelines. This stage provides direct visibility into communication bottlenecks, remote-operation overheads, and resource contention that arise in distributed quantum applications.

Finally, GsOQDC employs a custom DES and Monte Carlo framework to model the execution of remote quantum operations over the configured optical infrastructure. The simulation captures entanglement generation, distribution, storage, and consumption processes together with communication delays and hardware imperfections. Performance metrics, including entanglement fidelity, JET, communication overhead, entanglement-resource utilization, and waiting-time penalties caused by unavailable entangled resources, are reported in real time. By allowing attendees to modify network, hardware, and scheduling parameters and immediately observe their impact on application-level performance, the demonstration highlights the value of cross-layer design-space exploration for future OQDCs.

%------------------------------------------------------------------------
\section{Conclusions}
 
GsOQDC introduces a unified graphical framework for the design, scheduling, and evaluation of DQC over QDCs. By integrating optical-network modeling, distributed compilation, scheduling, and system-level DES-based simulation within a single environment, the platform enables end-to-end cross-layer analysis of how optical-network characteristics influence entanglement fidelity, communication overhead, entanglement-resource utilization, and JET. Through its user-friendly and cross-platform design, GsOQDC lowers the barrier to entry for DQC research while providing a practical platform for exploring the architectural and networking foundations of future OQDC.
 
%------------------------------------------------------------------------
\clearpage
\section*{Acknowledgements}
 
This work is supported by the Swedish Research Council (VR) and the UK EPSRC IQN Hub (EP/Z533208/1).
 
%------------------------------------------------------------------------

\printbibliography[]

@article{ionq2022,
  author    = {Christopher Monroe and Robert Raussendorf
               and Alex Ruthven and Kenneth R. Brown
               and Peter Maunz and L.-M. Duan
               and Jungsang Kim},
  title     = {Large-Scale Modular Quantum-Computer Architecture
               with Atomic Memory and Photonic Interconnects},
  journal   = {Physical Review A},
  volume    = {89},
  pages     = {022317},
  year      = {2014},
  doi       = {10.1103/PhysRevA.89.022317},
  url       = {https://doi.org/10.1103/PhysRevA.89.022317}
}

@article{main2025,
  author    = {D. Main and P. Drmota and D. P. Nadlinger
               and E. M. Ainley and A. Agrawal and B. C. Nichol
               and R. Srinivas and G. Araneda and D. M. Lucas},
  title     = {Distributed Quantum Computing across an
               Optical Network Link},
  journal   = {Nature},
  volume    = {638},
  pages     = {383--388},
  year      = {2025},
  doi       = {10.1038/s41586-024-08404-x},
  url       = {https://doi.org/10.1038/s41586-024-08404-x}
}

@article{caleffi2024,
  author    = {Marcello Caleffi and Michele Amoretti
               and Davide Ferrari and Jessica Illiano
               and Antonio Manzalini
               and Angela Sara Cacciapuoti},
  title     = {Distributed Quantum Computing: A Survey},
  journal   = {Computer Networks},
  volume    = {254},
  pages     = {110672},
  year      = {2024},
  doi       = {10.1016/j.comnet.2024.110672},
  url       = {https://doi.org/10.1016/j.comnet.2024.110672}
}

@misc{qiskit2024,
  author    = {Ali Javadi-Abhari and Matthew Treinish
               and Kevin Krsulich and Christopher J. Wood
               and Jake Lishman and Julien Gacon
               and Simon Martiel and Paul D. Nation
               and Lev S. Bishop and Andrew W. Cross
               and Blake R. Johnson and Jay M. Gambetta},
  title     = {Quantum Computing with {Qiskit}},
  year      = {2024},
  eprint    = {2405.08810},
  archiveprefix = {arXiv},
  url       = {https://arxiv.org/abs/2405.08810}
}

@article{tket,
  author    = {Seyon Sivarajah and Silas Dilkes
               and Alexander Cowtan and Will Sherburn
               and Alec Misra and Ross Duncan},
  title     = {\texttt{t|ket$\rangle$}: A Retargetable Compiler
               for {NISQ} Devices},
  journal   = {Quantum Science and Technology},
  volume    = {6},
  number    = {1},
  pages     = {014003},
  year      = {2020},
  doi       = {10.1088/2058-9565/ab8e92},
  url       = {https://doi.org/10.1088/2058-9565/ab8e92}
}

@misc{andres2024,
  author    = {Pablo Andres-Martinez and Daniel Mills
               and Timothy Forrer and Louis Henaut},
  title     = {{CQCL/pytket-dqc}: Distributed Quantum Computing
               Compiler Based on \texttt{pytket}},
  year      = {2024},
  url       = {https://github.com/CQCL/pytket-dqc}
}

@article{zhang2024gantt,
  author    = {Sen Zhang and Yipei Liu and Brian Mark
               and Weiwen Jiang and Zebo Yang and Lei Yang},
  title     = {Simulating Circuit Layout for Distributed
               Quantum Computing},
  journal   = {arXiv preprint},
  year      = {2024},
  eprint    = {2512.21403},
  archiveprefix = {arXiv},
  url       = {https://arxiv.org/abs/2512.21403}
}

@article{mirhosseini2020,
  author    = {Mohammad Mirhosseini and Alp Sipahigil
               and Mahmoud Kalaee and Oskar Painter},
  title     = {Quantum Transduction of Optical Photons from
               a Superconducting Qubit},
  journal   = {Nature},
  volume    = {588},
  pages     = {599--603},
  year      = {2020},
  doi       = {10.1038/s41586-020-3038-6},
  url       = {https://doi.org/10.1038/s41586-020-3038-6}
}

@misc{cisco_packet_tracer,
  author    = {{Cisco Systems}},
  title     = {Cisco Packet Tracer},
  year      = {2024},
  url       = {https://www.netacad.com/courses/packet-tracer}
}

@article{coopmans2021netsquid,
  author    = {Tim Coopmans and Robert Knegjens and Axel Dahlberg and
               David Maier and Loek Nijsten and Julio de Oliveira Filho and
               Martijn Papendrecht and Julian Rabbie and Filip Rozpedek and
               Matthew Skrzypczyk and Leon Wubben and Walter de Jong and
               Damian Podareanu and Ariana Torres-Knoop and
               David Elkouss and Stephanie Wehner},
  title     = {{NetSquid}, a {NET}work {S}imulator for {QU}antum
               {I}nformation using {D}iscrete events},
  journal   = {Communications Physics},
  volume    = {4},
  pages     = {164},
  year      = {2021},
  doi       = {10.1038/s42005-021-00647-8},
  eprint    = {2010.12535},
  archiveprefix = {arXiv}
}

@article{diadamo2021qunetsim,
  author    = {Stephen DiAdamo and Jakob N{\"o}tzel and Benjamin Zanger
               and Mehmet M. Bese},
  title     = {{QuNetSim}: A Software Framework for Quantum Networks},
  journal   = {IEEE Transactions on Quantum Engineering},
  volume    = {2},
  pages     = {1--12},
  year      = {2021},
  doi       = {10.1109/TQE.2021.3092395},
  eprint    = {2003.06397},
  archiveprefix = {arXiv}
}

@article{Krutyanskiy2023,
  author = {Krutyanskiy, V. and Galli, M. and Krcmarsky, V. and Baier, S. and Fioretto, D. A. and Pu, Y. and Mazloom, A. and Sekatski, P. and Canteri, M. and Teller, M. and Schupp, J. and Bate, J. and Meraner, M. and Sangouard, N. and Lanyon, B. P. and Northup, T. E.},
  title = {Entanglement of Trapped-Ion Qubits Separated by 230 Meters},
  journal = {Physical Review Letters},
  volume = {130},
  number = {5},
  pages = {050803},
  year = {2023},
  doi = {10.1103/PhysRevLett.130.050803},
  url = {https://link.aps.org/doi/10.1103/PhysRevLett.130.050803}
}

@article{Zhou2026PhotonicLink,
  author  = {Zhou, Yiyu and Wu, Yufeng and Li, Chunzhen and Shen, Mohan and Yang, Likai and Xie, Jiacheng and Tang, Hong X.},
  title   = {A 1-km photonic link connecting superconducting circuits in two dilution refrigerators},
  journal = {Nature Photonics},
  year    = {2026},
  doi     = {10.1038/s41566-026-01866-7}
}

@article{simdisq2025,
  author    = {Anonymous},
  title     = {{SimDisQ}: An End-to-End Distributed Quantum
               Circuit Simulator},
  journal   = {arXiv preprint},
  year      = {2025},
  eprint    = {2511.19791},
  archiveprefix = {arXiv},
  url       = {https://arxiv.org/abs/2511.19791}
}

@article{elyasi2025framework,
  author    = {Seyed Navid Elyasi and Paolo Monti and Jun Li and Rui Lin},
  title     = {A Framework for Quantum Data Center Emulation Using
               Digital Quantum Computers},
  journal   = {arXiv preprint},
  year      = {2025},
  eprint    = {2509.04029},
  archiveprefix = {arXiv},
  url       = {https://arxiv.org/abs/2509.04029}
}

\end{document}